\documentclass[12pt]{article}

\usepackage{graphicx} 
\usepackage[margin=1in]{geometry}
\usepackage{authblk}
\usepackage{color,soul}
\usepackage[table]{xcolor}
\definecolor{customcolor}{HTML}{94424F}
\definecolor{customBlue}{HTML}{5073b9}
\definecolor{customYellow}{HTML}{ffdc65}
\definecolor{customGreen}{HTML}{006868}
\definecolor{customCyan}{HTML}{00ffff}

\usepackage{colortbl}
\usepackage{tikz}
\usepackage[font=small,labelfont=bf]{caption}
\usepackage{subcaption}
\usepackage{cite}
\usepackage{amsmath}
\usepackage{amssymb}

\title{\textbf{\Large Bound State in the Continuum Supported Asymmetric Dome-shaped Dielectric Metasurface: Crossing and Avoided Crossing of Transmission with Applications}}

\author[$1,\dagger$]{\textcolor{customcolor}{\textbf{\small Ohidul Islam}}}
\author[$1,2,\dagger$]{\textcolor{customcolor}{\textbf{\small M. Hussayeen Khan Anik}}}
\author[1]{\textcolor{customcolor}{\textbf{\small Shakhawat Hossain Shakib}}}
\author[1]{\textcolor{customcolor}{\textbf{\small Nahid Hasan Niloy}}}
\author[1,3]{\textcolor{customcolor}{\textbf{\small Hriteshwar Talukder}}}
\author[$4,5, \star$]{\textcolor{customcolor}{\textbf{\small Shovasis Kumar Biswas}}}

\affil[1]{\scriptsize Department of Electrical and Electronic Engineering, Shahjalal University of Science and Technology, Sylhet 3114, Bangladesh.}
\affil[2]{Department of Electrical and Computer Engineering, University of Delaware, Newark, DE 19716, USA.}
\affil[3]{Department of Electrical and Computer Engineering, Boston University, Boston, MA, 02215, USA.}
\affil[4]{Department of Electrical and Computer Engineering, University of Wisconsin-Madison, Madison, WI, USA.}
\affil[5]{Department of Electrical and Electronic Engineering, Independent University, Bangladesh (IUB), Dhaka 1229, Bangladesh.}
\affil[$\dagger$]{These authors contributed equally to this work.}

\date{}

\begin{document}

\maketitle

\textbf{Abstract:} This work examines a symmetry-protected bound state in the continuum (BIC) supported unique all-dielectric dome-shaped metasurface (MS). The simple unit MS is made up of four dome-shaped nanobars of silicon in the top layer and silica glass as the substrate. Two strong transmission dips denoted as an electric dipole (ED) and magnetic dipole (MD) quasi-BIC with high Q-factor that surpasses the value of $\mathrm{10^4}$ are observed after the symmetry between the nanobars is broken by a small angle from their initial position. A crossover between the ED and MD resonances is noticed when the periodicity of the MS is changed in the y direction. In addition, transmission spectra show the avoided-crossing phenomenon of dipole quasi-BIC resonances when the superstrate's refractive index (RI) is reduced from its initial value. Other widely used dielectric materials are additionally employed in dome-shaped nanobars to evaluate their performances in terms of sharpness and near-zero transmission. Moreover, Other forms of symmetry breakdown such as diagonal width and length asymmetry have been investigated for their impact on the Q-factor. Finally, we have demonstrated two significant applications, including refractometric glucose sensing and third harmonic generation (THG). Our dome-shaped MS is roughly 300 times more efficient at generating third harmonics than a flat rectangular silicon film MS. Our proposed BIC and q-BIC-facilitated MS may provide a method for enhancing the functionality of biological sensors, multimodal lasing, optical switches, and nonlinear optics. 

\section{Introduction}
In view of the destructive interference of many modes of radiation, a bound state in the continuum (BIC) is a mathematical concept that refers to a non-expanding resonant state in an open framework that can't couple with radiating pathways propagating beyond the system \cite{rybin2017supercavity,hsu2016bound,joseph2021bound}. It is a confined state that interacts with an uninterrupted range of radiating waves that may transport energy with an extremely high-quality factor and can only exist for very high values of certain parameters or in perfect flawless never-ending structures \cite{monticone2014embedded,koshelev2018asymmetric,du2022dual}. In reality, a quasi-BIC may be achieved in which the Q factor becomes big yet finite near the BIC criterion in order to use BIC in nanophotonics \cite{wang2021all}. However, the presence of optical BIC modes is categorized as symmetry-protected BIC (SP-BIC), generated by the symmetry-constrained outcoupling, Fredrich-Wintegen (FW-BIC) or accidental BIC, and Fabry–Perot BIC (FP-BIC) \cite{hsu2013observation,friedrich1985interfering,marinica2008bound}. The SP-BICs may withstand minor structural flaws as long as the necessary symmetry is preserved. This symmetry can be broken by changing opto-geometrical parameters, in which case the BICs often shift to a resonant mode with a high Q-factor known as quasi-BIC (q-BIC) modes \cite{lepetit2014controlling}. To achieve q-BIC resonant phases with a high Q-factor, one is required to either slightly violate the excitation field symmetry with angled incidences or the in-plane/out-of-plane structural symmetry with normal incidence \cite{joseph2021bound,fan2019dynamic}. While out-of-plane symmetry breaking depends on changing the heights of a dimer or trimer cluster, in-plane symmetry breaking is often achieved by deforming the structure of the meta-atom or by creating a relative tiling between two meta-atoms \cite{kupriianov2019metasurface,salary2020tunable,kim2021high}.

\noindent BICs of various sorts have been used in innumerable photonic systems such as photonic crystals, metamaterials/metasurfaces, plasmonic structures, hybrid plasmonic-photonic structures, and fiber Bragg gratings \cite{lee2020bound,gao2019bound,gansch2016measurement,abujetas2019spectral}. In order to demonstrate the stimulation of high-Q resonances for the normal incidence of light, enhanced light focussing, wavefront and polarization regulating, and other applications, metasurfaces (MSs) are unique photonic frameworks with two-dimensional engineered periodic ensembles of subwavelength optical resonators \cite{koshelev2018asymmetric,zheludev2012metamaterials,zheng2021enhancing}. These metasurfaces may be quickly divided into two categories: plasmonic MS and all-dielectric MS. Plasmonic MS frequently interacts with metals in which significant quantities of light get captured to generate a considerable amount of heat as compared to photocarriers which can commonly degrade the performance \cite{tian2020high}. To counter the drawbacks of metallic MS, high-refractive-index (HRI) dielectric MSs have currently gained popularity owing to their benefits of minimal non-radiative losses and outrageous melting temperatures, with silicon being one of the viable HRI materials \cite{yang2018nonradiating}. Due to the significant scattering signals in Si-based dielectric nanostructure, including guided forward and backward scattering, electric dipole (ED) and magnetic dipole (MD) resonances have been examined \cite{terekhov2017multipolar,evlyukhin2011multipole}. In a Si-based MS, ED, and MD resonances can be generated to spectrally converge and oscillate in phase with one another without any re-emission of the electromagnetic (EM) field in the reverse direction, resulting in scattering cancellation in the reversed direction, known as the Kerker condition \cite{tian2018near,spinelli2012broadband,krasnok2012all}. By modifying certain parameters in MS, as well as the shapes and orientations of the dielectric materials on MS, it may be accomplished to control the overlap between ED and MD resonances, which can pave the way for a detrimental interference between the scattered field and the incident field in the forward direction, resulting in zero transmission in the spectrum \cite{tian2018near,yang2018nonradiating,odebo2017large}. Recent research has explored the use of these coupled-dipole compositions in MS with the help of BIC to improve sensing, EM-induced transparency, lasing, optical switching, filtering, and chirality \cite{abujetas2022tailoring,abujetas2021near}. BIC-supported MSs or dielectric nanoantennas, however, show intriguing applications in wave guiding, on-chip communications, beam steering, nonlinear harmonic production, photodetection, and even imaging \cite{joseph2021bound,grinblat2021nonlinear}.

\noindent Some new research has discovered that studies on BIC supported low or high-contrast dielectric gratings \cite{bulgakov2018avoided,huang2022tunable,joseph2021exploring}. Several investigations on plasmonic-photonic MSs that demonstrate critical coupling with BIC and qBIC formations have also been revealed \cite{ma2023tunable,kananen2022graphene,lu2020engineering}. All-dielectric MSs with BIC and qBIC support have become one of the most popular and extensively explored topics due to high Q factors in resonances, improved tunability and sensitivity, and a wide variety of applications that have already been discussed \cite{liu2022dirac,levanon2022angular,mohammadi2023active,grinblat2021nonlinear}. Some investigations on BIC-supported MS center their attention on ED and MD qBIC resonances, as well as their interactions with one another and the coupling of these resonances in homogeneous medium \cite{gao2022q,tian2020high,yu2021dielectric}. Although they are all groundbreaking in a sense, a significant number of the earlier research that has been brought out so far have either concentrated on an ordinary elliptical-shaped dielectric MS coupled with an in-depth BIC-related theoretical analysis or have concentrated on the applications of various MS nanostructures \cite{koshelev2018asymmetric,yu2021dielectric,tian2020high,anthur2020continuous,yu2022high,joseph2021bound}. Several investigations that focused on ED and MD qBIC resonances failed to achieve exceptionally high Q-factors, exhibit novel MS designs, or demonstrate potential applications in a computational context \cite{tian2020high,yang2018nonradiating,gao2022q,liu2018extreme}.

\noindent This work introduces an unconventional MS design consisting of four nanobars shaped like domes in a unit cell, where dipole coupling resonances for achieving zero transmission are demonstrated, along with two essential applications of BIC-supported MSs. Angle perturbation was introduced to break the symmetry of the MS which converts the the BIC intro qBIC modes.  When the periodicity of the MS is adjusted in the y direction, where zero transmission occurs, crossing between the ED-qBIC and MD-qBIC resonances is observed. Furthermore, avoided crossing between both resonances was detected when the period in the y-axis, P\textsubscript{y} was varied from 900 nm to 1060 nm and the refractive index value of the superstrate was reduced to 1.42 from its initial value. Additionally, we reported two vital applications of the MS including RI-based sensing of different glucose concentrations in a water-glucose solution and third harmonic generation with our proposed MS. Besides that, frequently employed dielectric materials (such as GaP, InP, and GaAs) are used in dome-shaped nanobars to assess their sharpness and near-zero transmission capability. Finally, we have investigated the Q-factor of the ED and MD qBIC resonances by exploring other methods of breaking symmetry in silicon nanobars such as diagonally breaking length and width symmetry of the nanobars.

\section{Methodology and Design}
\subsection{Theoretical Model}
Our proposed MS consists of dome-shaped elements to evaluate the BIC mode can be analyzed by employing the coupled mode theory. A system can be considered for two resonances (ED-qBIC and MD-qBIC) coupled with two ports which can be expressed by the equation below \cite{gao2022q}. 

\begin{equation}
    \left[\begin{array}{l}
        a_1 \\
        a_2
    \end{array}\right]=j\left[\begin{array}{cc}
\omega_{01}+j \gamma_1 & k+j \gamma_{12} \\
k+j \gamma_{21} & \omega_{02}+j \gamma_2
\end{array}\right]\left[\begin{array}{l}
a_1 \\
a_2
\end{array}\right]+\left[\begin{array}{ll}
k_{11} & k_{12} \\
k_{21} & k_{22}
\end{array}\right]\left[\begin{array}{l}
s_{1+} \\
s_{2+}
\end{array}\right]
\end{equation}

\noindent Here, where the $\mathrm{[a1, a2]^T}$ represent the time-dependent amplitudes of the ED-qBIC and MD-qBIC resonances, and $\mathrm{\omega_{01}}$ and $\mathrm{\omega_{02}}$ are their resonant frequencies, respectively. Also, k is the direct coupling rate of these two resonances. The $\mathrm{\gamma_1}$ and $\mathrm{\gamma_2}$ refer to the radiative loss of the system thus it is directly related to the radiative Q-factors($\mathrm{Q_R}$) by this equation $\mathrm{\gamma_n = \frac{\omega_{0n}}{2Q_{Rn}}}$. Due to the symmetry of the system, $\mathrm{\gamma_{12}=\gamma_{21}=\sqrt{\gamma_1\gamma_2}}$.  The $\mathrm{k_{ij}}$ is the coupling coefficient between the mode i and the port j ($\mathrm{i,j \in {1,2}}$). The $\mathrm{\left [s_{1+}, s_{2+} \right]^T}$ and $\mathrm{\left[s_{1-}, s_{2-} \right]^T}$ denotes the input and output wave amplitude of the excited resonance modes at the port 1 and 2, respectively as demonstrated in Eqn. (2). 

\begin{equation}
    \left[\begin{array}{l}
        s_{1-} \\
        s_{2-}
    \end{array}\right]=\left[\begin{array}{ll}
    r_d & t_d \\
    t_d & r_d
    \end{array}\right]\left[\begin{array}{l}
    s_{1+} \\
    s_{2+}
    \end{array}\right]+\left[\begin{array}{ll}
    d_{11} & d_{12} \\
    d_{21} & d_{22}
    \end{array}\right]\left[\begin{array}{l}
        a_1 \\
        a_2
    \end{array}\right]
\end{equation}

\noindent Here, $r_d$ and $t_d$ are the direct reflection and transmission coefficients between the ports in the absence of the resonant modes, and $d_{ij}$ is the coupling coefficient between the port j and the mode i. When the input wave is incident from port 1, the transmission coefficient from port 1 to port 2 may be calculated by $t_{21}(\omega)=\frac{s_{2-}}{s_{1+}}$. The coupling of qBICs is extensively explored in the result section, with particular emphasis on coupling coefficients.

\noindent We also analyze the proposed metasurface structure for nonlinear harmonic generation. The nonlinear optical interaction for a silicon medium is caused by induced polarization which is governed by the equation below \cite{boydbook}.

\begin{equation}
    \tilde{P}^{(3)}(t) = \varepsilon_0 \left[\Tilde{\chi}^{(3)}E^3(t) \right]
    \label{eqn:non_polarization}
\end{equation}

\noindent Here, $\varepsilon_0$ is the vacuum permittivity, E(t) is the strength of the electric field, and $\mathrm{\Tilde{\chi}^{(3)}}$ is the third-order nonlinear optical susceptibility tensor. we considered a diagonal anisotropy tensor for $\mathrm{\Tilde{\chi}^{(3)}}$ with a  value of $\mathrm{2.45 \times 10^{-19}}$ $\mathrm{m^2V^{-2}}$ \cite{hahnel2023multi}. The third harmonic generation (THG) process is associated with the annihilation of three photons of frequency and the subsequent creation of a single photon with three times the frequency. When light is pumped on the surface of the structure, its electromagnetic field excites the unbound electrons which leads them to vibrate in their ionic core. This oscillation introduces a nonlinear shift that leads to an anharmonic response to the electron's motion in relation to the applied electric field \cite{fomichev2010linear}.

\subsection{Metasurface Structure}
\begin{figure*}[ht!]
    \centering
    \captionsetup{width=0.95\linewidth}
    \includegraphics[width=14cm]{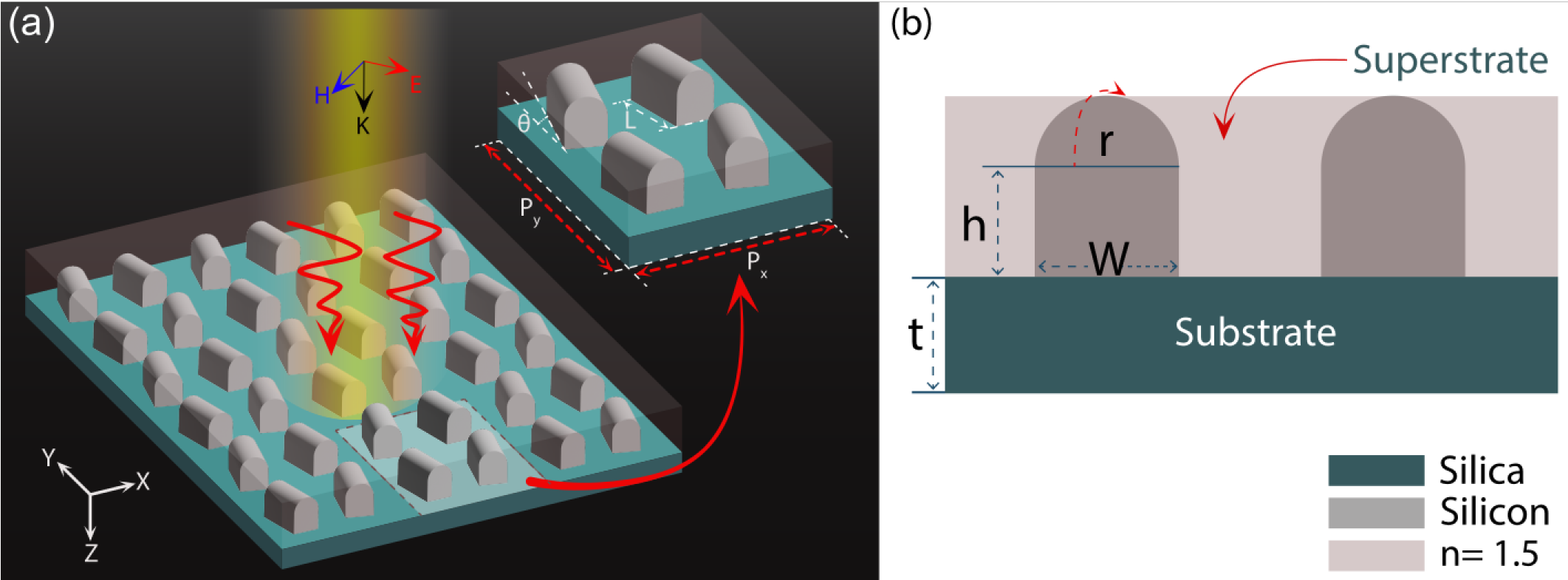}
    
    \caption{(a) Schematic of the all-dielectric metasurface with the additional indication of a unit cell with P\textsubscript{x} =  P\textsubscript{y} = 860 nm, $ \mathrm{\theta} = 9^0$ and L = 350 nm. The unit cell consists of four dome-shaped Silicon (Si) nanobars. It exhibits in-plane symmetry in the x-y plane and  $ \mathrm{\theta}$ was used to break the symmetry of the metasurface. (b) A 2-D representation of the unit cell where the Silica ($\mathrm{SiO_2}$) has been used as the substrate that has a thickness (t) = 220 nm and a material of refractive index, n = 1.5 was used as the superstate.}
    \label{fig:design}
\end{figure*}

To support symmetry-protected BIC, we proposed a periodic asymmetric dome-shaped all-dielectric metasurface with in-plane symmetry. Figure \ref{fig:design}(a) and Fig. \ref{fig:design}(b) illustrate the conceptual 2-D and 3-D views of the structure. The metasurface has a two-layer structure consisting of a periodic array of silicon (Si) domes atop a silica ($\mathrm{SiO_2}$) substrate. Values of refractive index for both the materials with discussion of simulation environment are briefly analyzed in \textcolor{blue}{supplement document 1}. In comparison to other materials, silicon structures are considerably more advantageous due to their high transmission rate \cite{MALEKFAR2018140, Ruihua_2021}, minimal losses \cite{wang2021bifunctional, Li:21}, and well-established fabrication method \cite{Gylfason2012, Cho2013}.  Silica, on the other hand, functions as the substrate for this structure because of its transparency in the visible and infrared range enabling efficient light transmission. This facilitates simple light manipulation and controls \cite{Almeida2004}.

\noindent In order to simulate the structure using the finite difference time domain method (FDTD), we considered a unit cell with a period of P\textsubscript{x} = P\textsubscript{y} = 860 nm in the x and y direction and the substrate thickness, t = 220 nm. Four domes are placed on top of a unit cell. These domes are placed inside the superstate, a material of refractive index, n = 1.5. Each dome has properties of h = 175 nm, r = 70 nm, and $\mathrm{W = 2 \times r}$ = 140 nm. The orientation of each dome is characterized by a rotation angle of $ \mathrm{\theta} = 9^0$. This design may initially appear difficult to fabricate. Nevertheless, modern fabrication techniques have advanced significantly, and even more complex designs have been fabricated in the past \cite{LI2020100584, YeungTsaiKingPhamHoLiangKnightRaman+2021+1133+1143, Wang19, LaSpada2019}.  As depicted in Fig. \ref{fig:design}(a), a normal incident plane wave propagated along the z-axis when the electric field, E was y-polarized. Since the ideal symmetry-protected BIC may be converted into the quasi-BIC with high Q factor resonances, we incorporated a rotation angle ($\mathrm{\theta}$) as in-plane symmetry-breaking perturbations to study the characteristics of the BIC or quasi-BIC. Prior to analyzing the outcomes of our metasurface, we conducted a simulation verification process. This involved comparing our simulation results, obtained by reproducing a design described in a prior study, with the experimental findings provided in that study which is discussed in \textcolor{blue}{supplement document 1}. 

\section{Results Discussion}
\subsection{Formation of BIC and qBIC with Field Distribution}
\begin{figure*}[ht!]
    \centering
    \captionsetup{width=0.95\linewidth}
    \includegraphics[width=14cm]{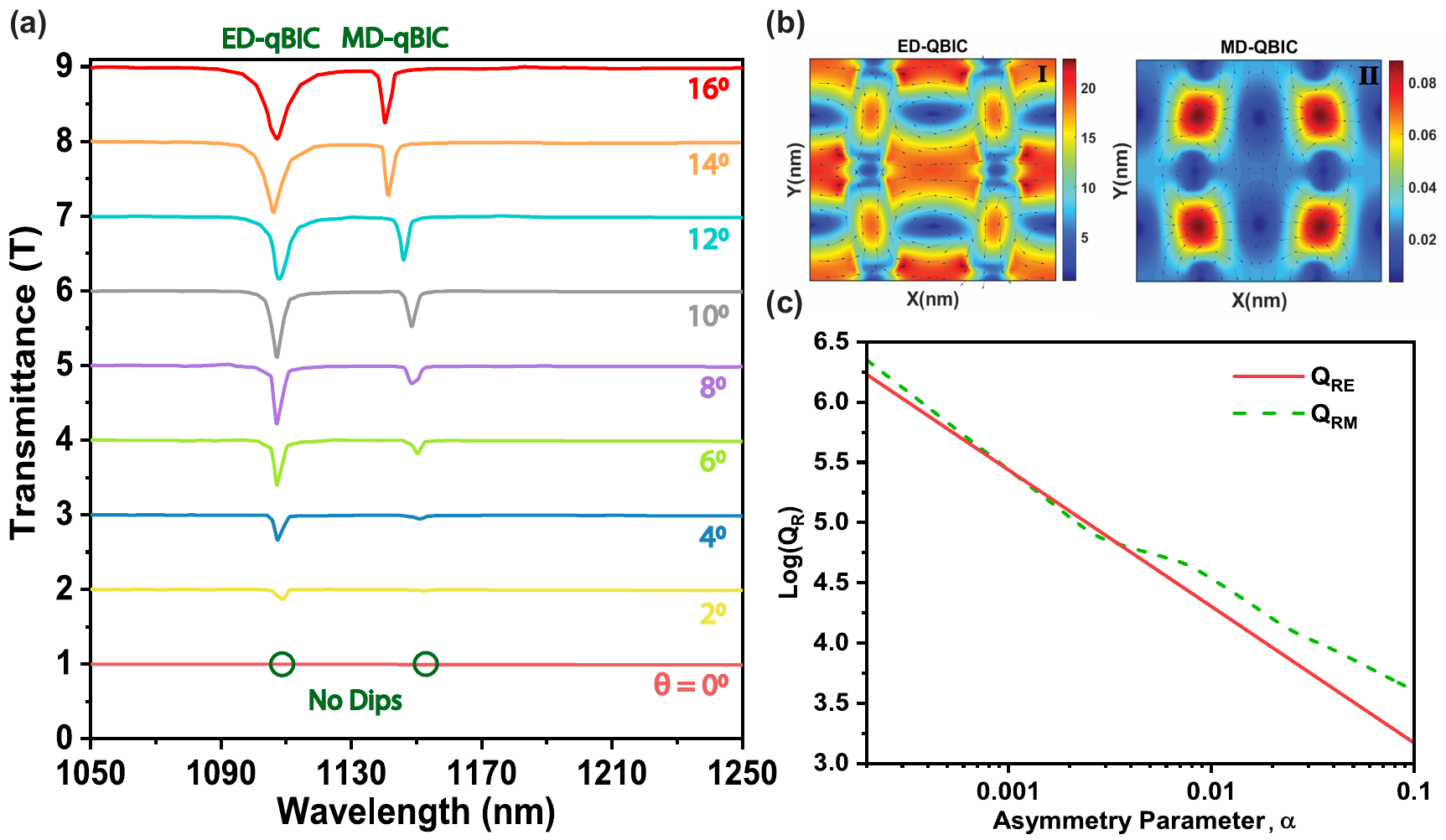}
    
    \caption{(a) Transmission spectrum of the MS depicting the impact of different rotation angles ($\mathrm{\theta}$) of the domes. Introducing non-zero values to the $\theta$ causes two dips in the transmission curve corresponding to ED-qBIC and MD-qBIC, respectively. At $\mathrm{\theta = 9^0}$, the dips are located at 1110.1 nm and at 1154.28 nm. (b) - \MakeUppercase{\romannumeral1}  Electric field distribution of the unit cell for $\mathrm{\theta = 9^0}$  at 1110.11 nm as the ED-qBIC resonance happened at this wavelength. (b) - \MakeUppercase{\romannumeral2} Magnetic field distribution of the unit cell for $\mathrm{\theta = 0^0}$ at 1154.28 nm as the MD-qBIC resonance happened at this wavelength. Vector fields are denoted by the blue arrow lines for both figures.  (c) Radiative Q-factor in logarithm scale as a function of angle asymmetry parameter ($\alpha$) of the metasurface where QRE and QRM represent Q-factor for ED-
    qBIC and MD-qBIC resonance, respectively.}
    \label{fig:BIC}
\end{figure*}

Bound states in the continuum (BIC) are characterized by destructive interference, which occurs when the coupling constants with all radiating waves disappear by accident as a result of continuous adjustment of parameters. Figure \ref{fig:BIC}(a) depicts the transmission spectra of the dome-shaped MS for different values of rotational angle, $\mathrm{\theta}$. When the $ \mathrm{\theta} = 0^0$ which means that the MS is maintaining the in-plane symmetry, there is no spectral line width that can be noticed in the transmission curve. This indicates that the Q factor is infinite, which corresponds to the formation of BIC modes.  Changing the geometric characteristics of the system is necessary in order to transition from the BIC state to the qBIC state. This enables the extraction of energy from the BIC state while maintaining a long-lived state with narrow linewidths, and high q-factor values. So as to disrupt the in-plane symmetry, the angle of the domes was altered from 0$^0$ to 16$^0$ to analyze the transmission behavior of the quasi-BIC as a function of $ \mathrm{\theta}$.

\noindent Transmission spectra in Fig. \ref{fig:BIC}(a) display two dips at 1110.10 nm and 1154.28 nm corresponding to ED-qBIC and MD-qBIC resonances, respectively. Both qBIC resonances are susceptible to angle perturbations, as evidenced by the increasing linewidth and increasing dips with increasing $\theta$. In contrast to the situation in which $ \mathrm{\theta} = 0^0$, the net ED-qBIC and MD-qBIC resonances are now capable of being excited by a y-polarized incident plane wave and of coupling to free space radiation due to the fact that the symmetry protection has been broken. Figure \ref{fig:BIC}(b) shows an illustration of the field distributions at 1110.1 nm and 1154.2 nm corresponding to the ED-qBIC and MD-qBIC resonances for $\mathrm{\theta=9^0}$ in the x-y plane inside one unit cell. These field distributions reveal that both dipole quasi-BIC resonances display reasonably high field intensity enhancement. Both the electric and magnetic dipole moments that are generated are dissociated and display symmetry in the opposite direction around the plane of the mirror. When viewed in relation to the x-y plane, the moments of electric dipoles that oscillate along the x-axis are equal and symmetrical. Therefore, magnetic dipole moments in the x-z plane that form a displacement current loop are antisymmetric (odd) in comparison to the x-y plane.

\noindent Figure \ref{fig:BIC}(c) shows the radiative Q factor in log scale of the dome-shaped MS where Q\textsubscript{RE} and Q\textsubscript{EM} indicate the radiative Q factor of ED-qBIC and MD-qBIC resonance, respectively. For our suggested configuration where $\mathrm{\theta = 9^0}$, the values for Q-factor are 10680.7 for Q\textsubscript{RE} and 12410 for Q\textsubscript{EM}. Increases in the $ \mathrm{\theta} $ result in a reduction in the radiative Q-factors of the ED-qBIC and MD-qBIC resonances and a widening of the spectral lines caused by their coupling to free space. The Q\textsubscript{R} can be related to in such $\mathrm{Q_R \propto \alpha^{-2}}$ where $\alpha$ is the asymmetry parameter defined by $\mathrm{\alpha = sin(\theta)}$. Increasing the value of the asymmetry parameter induces a non-zero dipole moment, which in turn transforms the BIC into the accessible q-BIC.

\subsection{Crossing and Anti-crossing of Transmission}
\begin{figure*}[ht!]
    \centering
    \captionsetup{width=0.95\linewidth}
    \includegraphics[width=14cm]{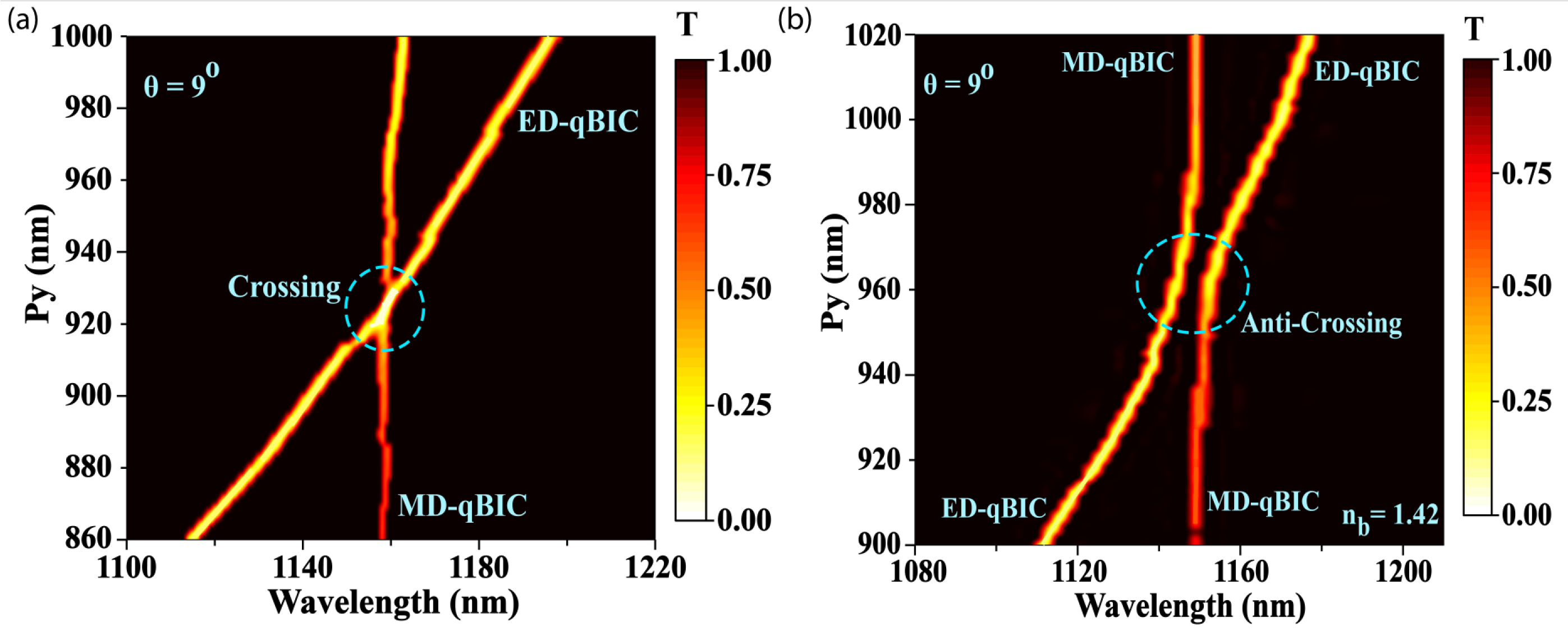}

    \caption{(a) Transmission spectrum of the MS at $\mathrm{\theta = 9^0}$ as a function of period $\mathrm{P_y}$ shows a crossing of the ED-qBIC and MD-qBIC for $\mathrm{P_y}$ = 925 nm at 1152 nm wavelength. (b) Transmission spectrum of the MS when the superstate's refractive index was changed from 1.5 to 1.42. Avoided Crossing of ED-qBIC and MD-qBIC was observed at $P_y = 960$ nm. Both crossing and anti-crossing were marked by a small circle.}
    \label{fig:crossing-anticrossing}
\end{figure*}

 The progression of transmission spectra for both ED-qBIC and MD-qBIC with respect to $\mathrm{P_y}$ at $\mathrm{\theta = 9^0}$ is illustrated in Fig. \ref{fig:crossing-anticrossing}. It can be seen that even though MD-qBIC resonance wavelength remained nearly identical throughout the entire spectra, ED-qBIC resonance wavelength shifts to the right side of the wavelength as the P\textsubscript{y} increases, getting closer to MD-qBIC resonance. At P\textsubscript{y} = 925 nm, both resonances overlap with each other as indicated by the circle in Fig. \ref{fig:crossing-anticrossing}(a). The crossings area confirms that the coupling coefficient, k is zero, proving that both resonances are orthogonal. It is consistent with the concept of an extreme Huygens' metasurface because the transmittance approaches unity with a large quality factor, Q in the crossing location \cite{zhao2020terahertz,wang2021bifunctional,liu2018extreme}.         
However, when the superstate is replaced by a refractive index, $\mathrm{n = 1.42}$ material, the vertical symmetry between the silica substrate and the superstate is broken. This causes a nonzero value of the coupling coefficient.  In this case, the metasurface exhibits omega-type bianisotropy which leads to coupling between MD-qBIC and ED-qBIC resonances  where both modes strongly exchange energy \cite{albooyeh2011substrate}. At $\mathrm{\theta = 9^0}$, varying $\mathrm{P_y}$ from 1080 nm to 1220 nm, the corresponding outcome of this modification can be visualized in Fig. \ref{fig:crossing-anticrossing}(b) where instead of overlapping, an avoided crossing was observed for $\mathrm{P_y} = 960$ nm. The \textcolor{blue}{supplement document 1} provides a comprehensive depiction of the transmission curves pertaining to both the crossing and anti-crossing phenomena seen in the interaction between the resonances. 

\subsection{Influence of Different Dielectric Materials}
We attempted to evaluate the performance of other commonly used materials (GaAs, InP, and GaP) for this structure by incorporating them into different dome components. A comparison has been made in terms of crossing wavelength, crossing $\mathrm{P_y}$, lowest transmission, and FWHM (Full Width at Half Maximum) of the qBIC resonance. Table  \ref{tab:diff_mats} demonstrates that although all material combinations have about similar crossing wavelength and crossing $\mathrm{P_y}$, the FWHM for the Si-GaAs, GaAs-Si, and InP-InP combinations is substantially larger with values of 4.101 nm, 5.21 nm, and 4.36 nm, respectively. Since we are aware that a larger FWHM leads to a lower Q-factor \cite{kupriianov2019metasurface}, only Si-Si and Si-GaP pairings with FWHM values of 2.123 nm and 1.61 nm, respectively, ought to be considered when constructing an MS with a high Q-factor value. Although the FWHM of the Si-Si combination is greater than that of the Si-GaP combination, the Si-Si combination allows for the lowest T (0.001) to be obtained, making it the more overpowering material combination for this MS. The \textcolor{blue}{supplement document 1} contains visual representations of the transmission curves corresponding to the various material implementations in the dielectric nanobars.

\renewcommand{\arraystretch}{1.1}
\begin{table}[!ht]
    \centering
    \scriptsize
    \caption{Comparative analysis of inserting different dielectric materials.}
    
    \begin{tabular}{>{\centering\arraybackslash}m{4cm} >{\centering\arraybackslash}m{3cm} >{\centering\arraybackslash}m{2cm} >{\centering\arraybackslash}m{1.9cm} >{\centering\arraybackslash}m{1.9cm} }
    \rowcolor{white}\arrayrulecolor{black}
    \hline
       \textbf{Dome Structure (upper half circle \& lower rectangle)}  & \textbf{Crossing Wavelength (nm)} & \textbf{Crossing $P_y$ (nm)} & \textbf{Lowest T} & \textbf{FWHM (nm)} \\
    \hline
    \vspace{2pt}
    \includegraphics[width=9mm]{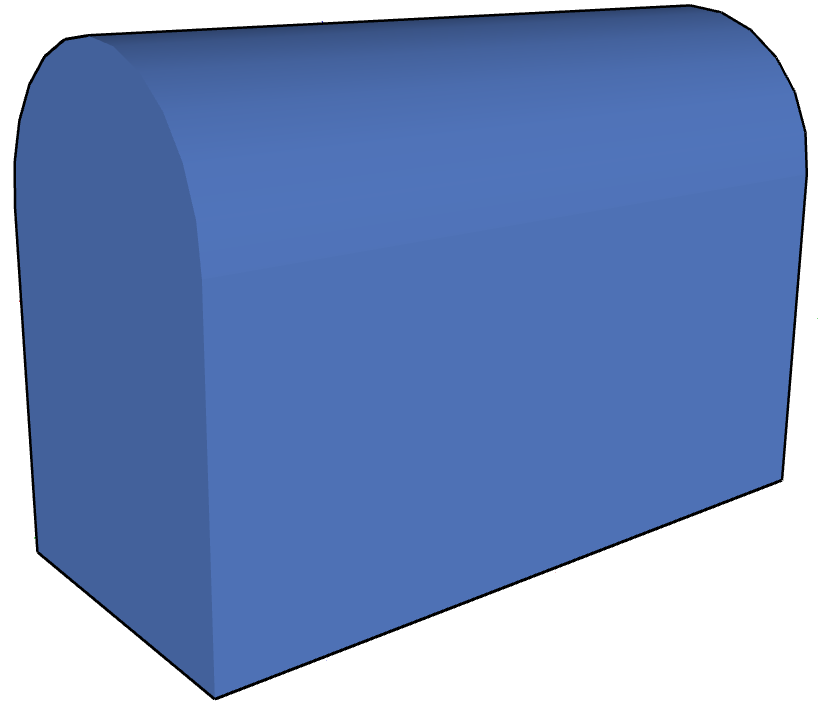}
     & 1150.4 & 925 & 0.001 & 2.123 \\ 
    \includegraphics[width=9mm]{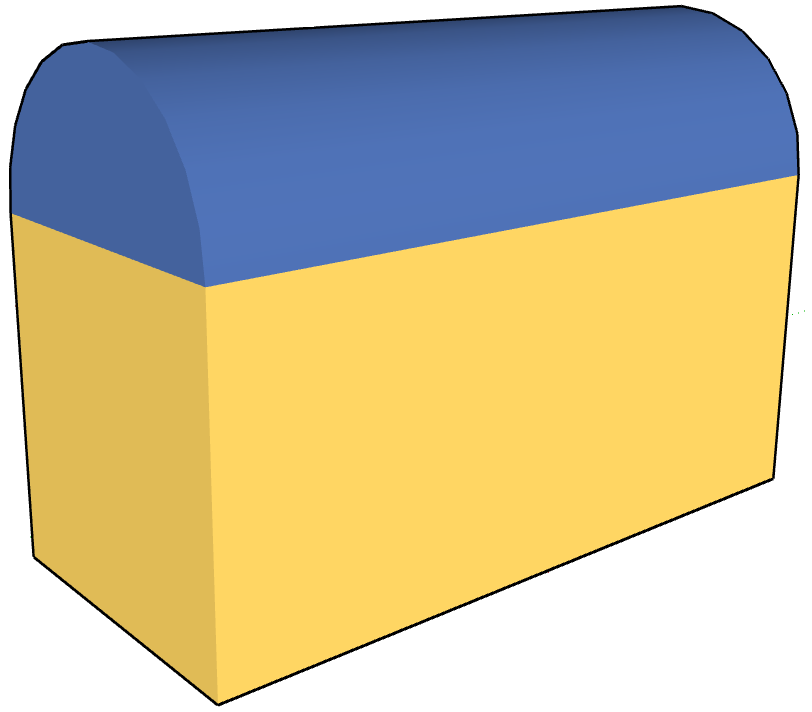} & 1142.3 & 920 & 0.009 & 4.101 \\
    \includegraphics[width=9mm]{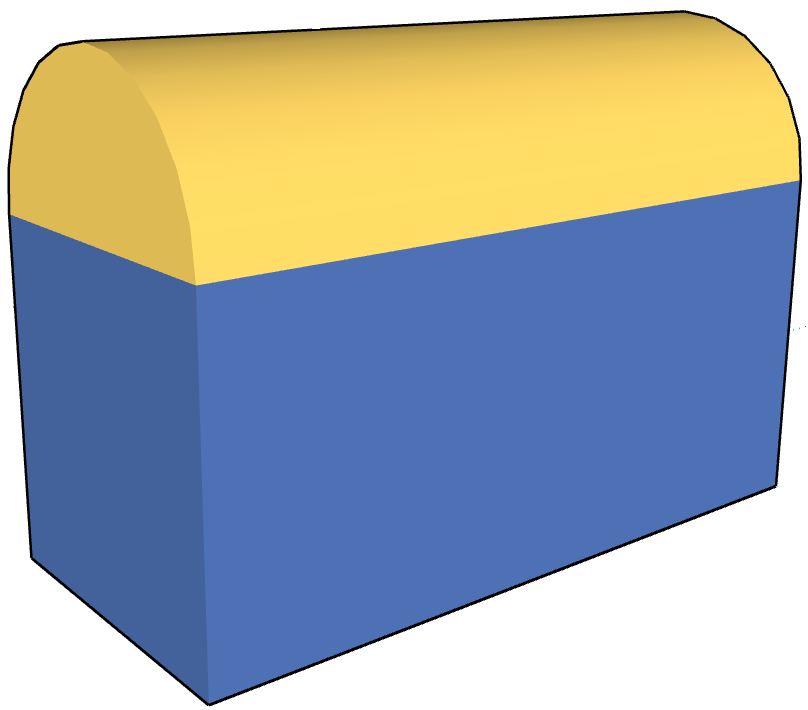} & 1146.3 & 920 & 0.207 & 5.21 \\
    \includegraphics[width=9mm]{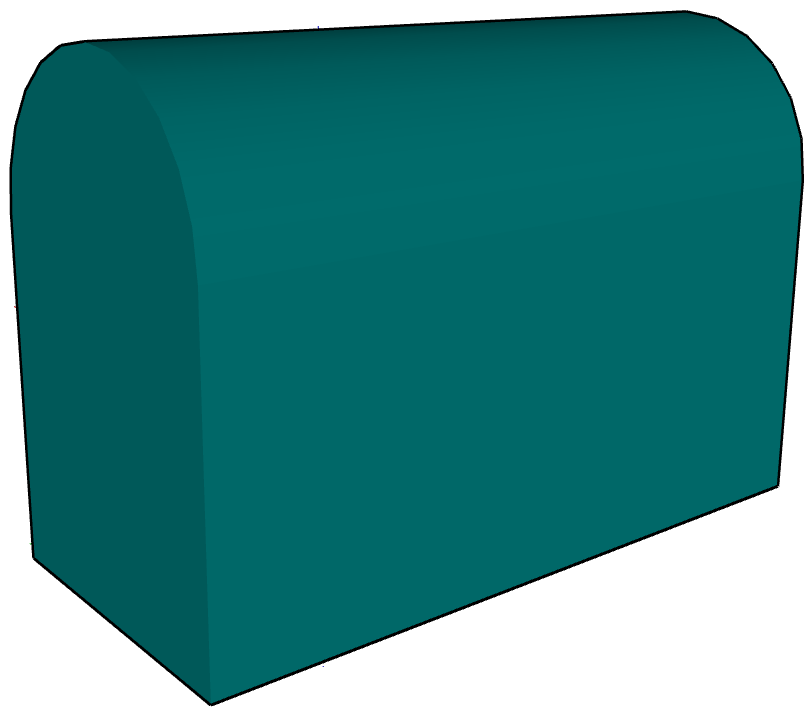} & 1103.3 & 897 & 0.206 & 4.36 \\
    \includegraphics[width=9mm]{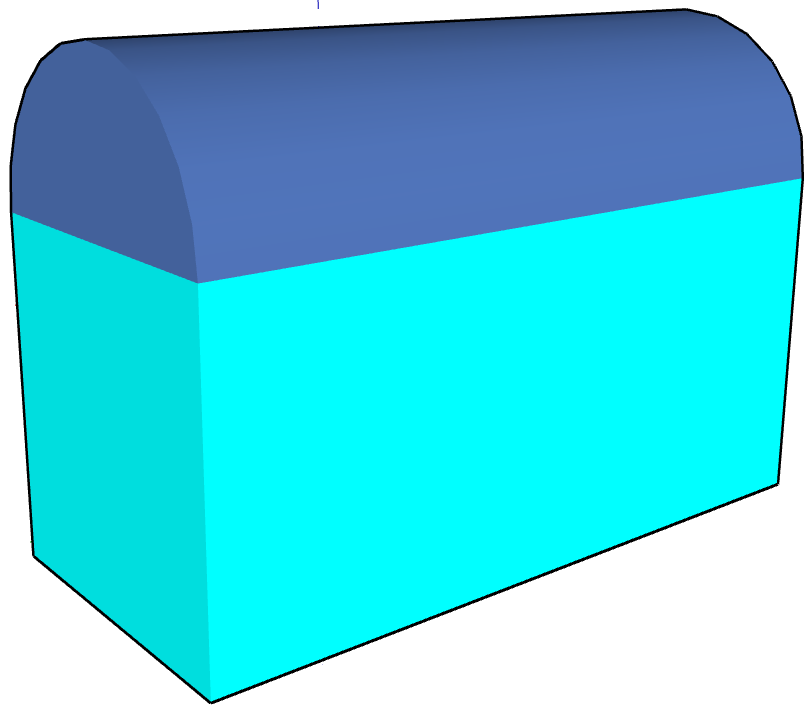} & 1097.9 & 900 & 0.07 & 1.61\\
    \hline 
    \textbf{Label} & 
\vspace{4pt}
    \tikz \fill[customBlue] (0,0) rectangle (0.7,0.3); Si
 & 
\vspace{4pt}
    \tikz \fill[customYellow] (0,0) rectangle (0.7,0.3); GaAs
 & 
\vspace{4pt}
    \tikz \fill[customGreen] (0,0) rectangle (0.7,0.3); InP
 & 
\vspace{4pt}
    \tikz \fill[customCyan] (0,0) rectangle (0.7,0.3); GaP\\

\hline

\centering
    \end{tabular}
    
    \label{tab:diff_mats}
\end{table}

\subsection{Applications}
\subsubsection{Third Harmonic Generation}
\begin{figure*}[ht!]
    \centering
    \captionsetup{width=0.95\linewidth}
    \includegraphics[width=14cm]{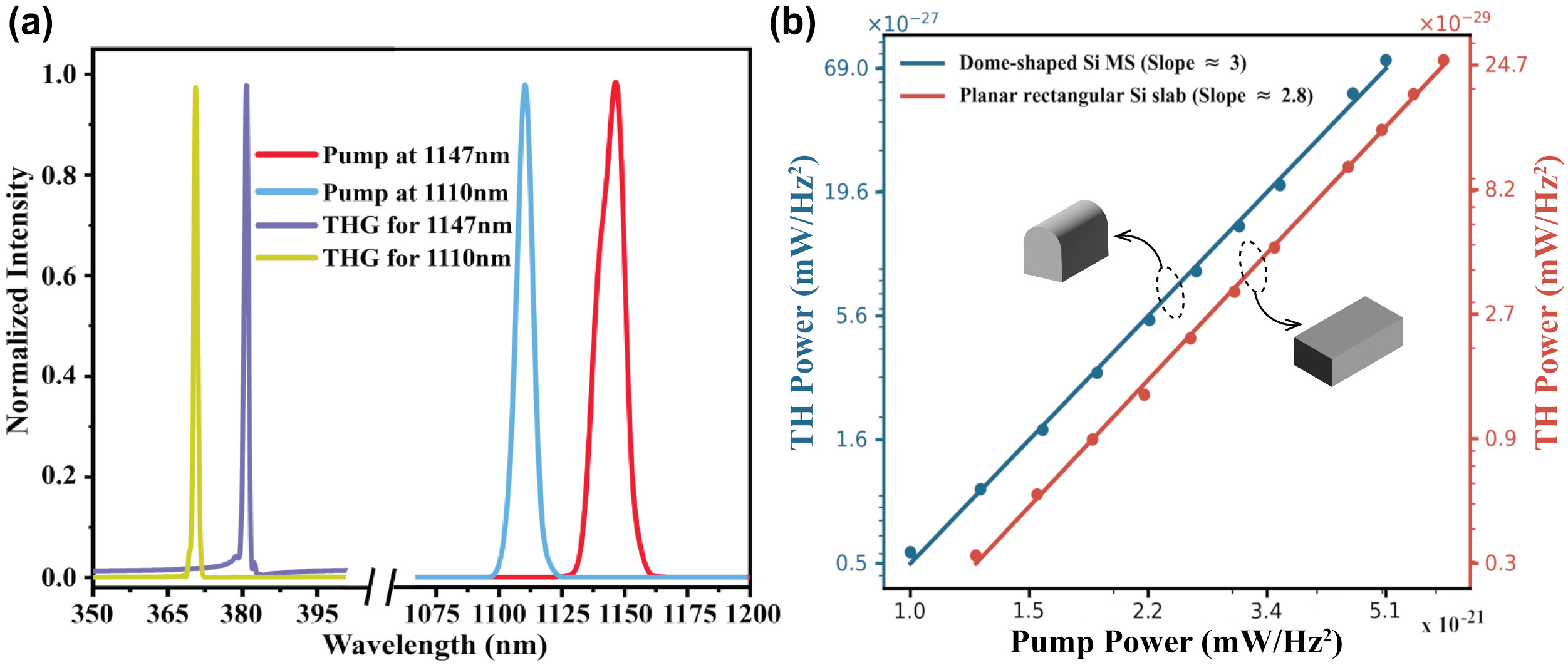}
    \caption{(a) Normalized intensity spectra of plane waves which were pumped at 1110 nm and 1147 nm as well as generated third harmonics at 370.46 nm and 381.15 nm, respectively for those pump spectra. (b) TH efficiency as a function of pump power for our proposed MS and an MS comprised of rectangular elements, with all other parameters remaining unchanged.}
\label{fig:harmonic}
\end{figure*}

Figure \ref{fig:harmonic}(a) depicts the normalized THG outcome for the metasurface structure together with the normalized intensity of the pump. We pumped a plane wave into the structure with a 2.2 THz bandwidth at 1110 nm and 1147 nm near the ED-qBIC and MD-qBIC resonances, respectively, and observed peaks in intensity at 370.46 nm and 381.15 nm, respectively, in the electric field spectra which clearly indicates the generation of the third harmonic in both cases. The \textcolor{blue}{supplement document 1} delivers a brief analysis of nonlinear simulation. The THG efficiency was calculated using the $\mathrm{\eta_{THG} = \frac{P_{TH}}{P_{Pump}}}$ equation. In order to determine the pump and TH power, the Poyting vector was integrated over the x-y plane. In terms of \mbox{$\mathrm{P_{TH}}$} as a function of pump power in logarithm scale, figure \ref{fig:harmonic}(b) illustrates a comparison between our proposed metasurface and a metasurface consisting of rectangular elements while maintaining all other parameters the same as our proposed metasurface.

\noindent P\textsubscript{TH} follows a third-order power law dependence on the pump power. Performing a curve-fitting on the data from figure \ref{fig:harmonic}(b), we found the slope of the dome-shaped Si metasurface and the planar rectangular Si slab metasurface is roughly 3 and 2.8, respectively. Our computation yielded a maximum THG efficiency of $\mathrm{1.48 \times 10^{-5}}$ for the dome-shaped metasurface, whereas a flat rectangular Si slab metasurface achieved a maximum efficiency of \mbox{$\mathrm{4.19 \times 10^{-8}}$}. That indicates that, in comparison to the planar rectangular Si slab metasurface, our suggested dome-shaped Si metasurface exhibits 353.22 times higher THG efficiency. For references with other published works involving THG in Si metasurface, Shumei et al. conducted a study where they reported a maximum THG efficiency of \mbox{$\mathrm{1.76 \times 10^{-7}}$} in a Si metasurface composed of a 2D periodic array of nanoapertures \mbox{\cite{chen2018third}}. Another research by Ze et al. demonstrated THG with maximum efficiency of \mbox{$\mathrm{3.6 \times 10^{-6}}$} with resonant Si membrane metasurface \mbox{\cite{zheng2023third}}.  We can see a significant improvement in THG efficiency utilizing dome-shaped metasurface in comparison to these published works.

\subsubsection{Refractive Index (RI) based Sensing}
The confined light of the qBIC modes has an evanescent tail that usually propagates into the substrate medium, and the interaction and any modifications in the substrate medium have a significant effect on the resonance properties \cite{sahoo2017high}.  
\begin{figure*}[ht!]
    \centering
    \captionsetup{width=0.95\linewidth}
    \includegraphics[width= 13cm]{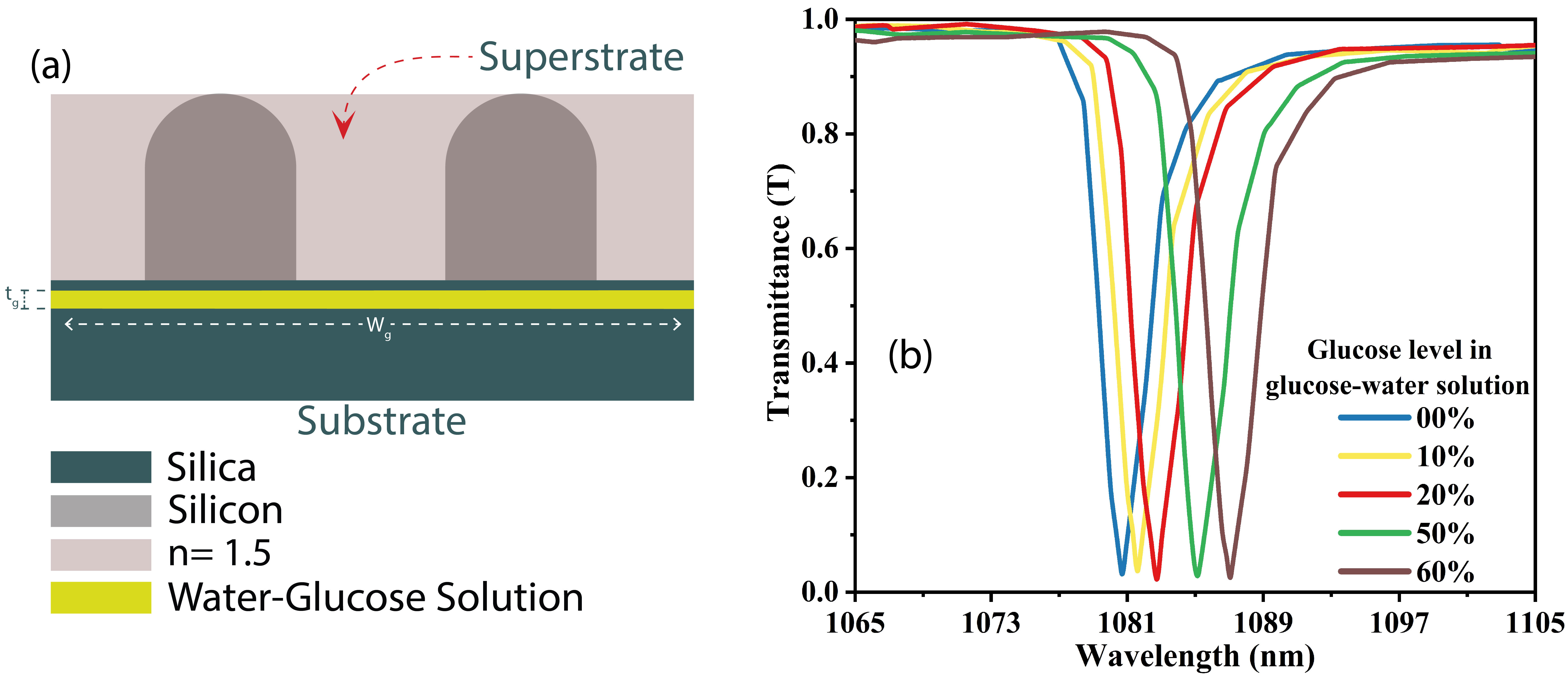} 
            
    \caption{(a) 2D view of the unit cell depicting the formation of a chamber on the upper surface of the substrate to accommodate the dielectric analyte. The yellow rectangle denotes the chamber filled with water-glucose solution. The chamber has properties of thickness, t\textsubscript{g} = 110 nm, width, W\textsubscript{g} = 840 nm, and length = 840 nm. (b) Transmission spectra of the metasurface when introducing different glucose levels (00\%, 10\%, 20\%, 50\%, and 60\%) in glucose-water solution as the sensing material.}
    \label{fig:sensing}
\end{figure*}
The addition of a target analyte to the substrate can alter the dielectric environment.  Alteration in the local dielectric composition can be detected by observing shifts in the transmission or reflection curve at resonant wavelengths. In this study, we investigated whether the proposed metasurface can detect various glucose concentrations in a glucose-water solution. As shown in figure \ref{fig:sensing}(a), a hollow chamber with dimensions of 860 nm $\times$ 860 nm $\times$ 110 nm was established at the top of the substrate to insert and support the analyte. Then, one by one 00\%, 10\%, 20\%, 50\%, and 60\% concentration of glucose-water solution was incorporated into the system. It can be seen from the figure \ref{fig:sensing}(b) that as the glucose level increases, the resonant wavelength shifts to longer wavelengths, exhibiting a distinct peak at 1078 nm, 1081.50 nm, 1083 nm, 1085 nm, and 1087.5 nm, respectively, for the various solutions. These outcomes validate the capability of RI sensing of the MS. However, Wavelength sensitivity (WS) refers to the change in resonant transmission peaks that occurs when the refractive index of the analyte is altered. In order to comprehend the sensor's exceptional detecting capabilities and its ability to detect even the smallest quantities, it is necessary to consider the WS of the metasurface. 
Nevertheless, the figure of merit (FOM) is another essential indicator for determining the sensor's detection limit. The ratio of the WS to the Gaussian transmission curve's full-width half maximum (FWHM) is known as the FOM. Because it makes it easier to interpret the output response curve from undesired noise signals, a larger FOM value is beneficial. 
WS can be calculated as $S_\lambda = \frac{\Delta\lambda_{peak}}{\Delta n_a}$ whereas FoM can be obtained from: $FoM=\frac{S_{\lambda}}{FWHM}$.
The refractive indices of glucose level in glucose-water solution are 1.33, 1.348, 1.364,1.42 and 1.442 for percent glucose of 0\%, 10\%, 20\%, 50\% and 60\%, respectively. Calculating the corresponding resonance shift in wavelength, we have found the highest WS to be 85.68 nm/RIU while changing the glucose level from 50\% to 60\%. For this transition in the glucose level, we have calculated the FoM to be 25.13 $RIU^{-1}$.

\subsection{Asymmetry Analysis}
\begin{figure*}[ht!]
    \centering
    \captionsetup{width=0.95\linewidth}
    \includegraphics[width=14cm]{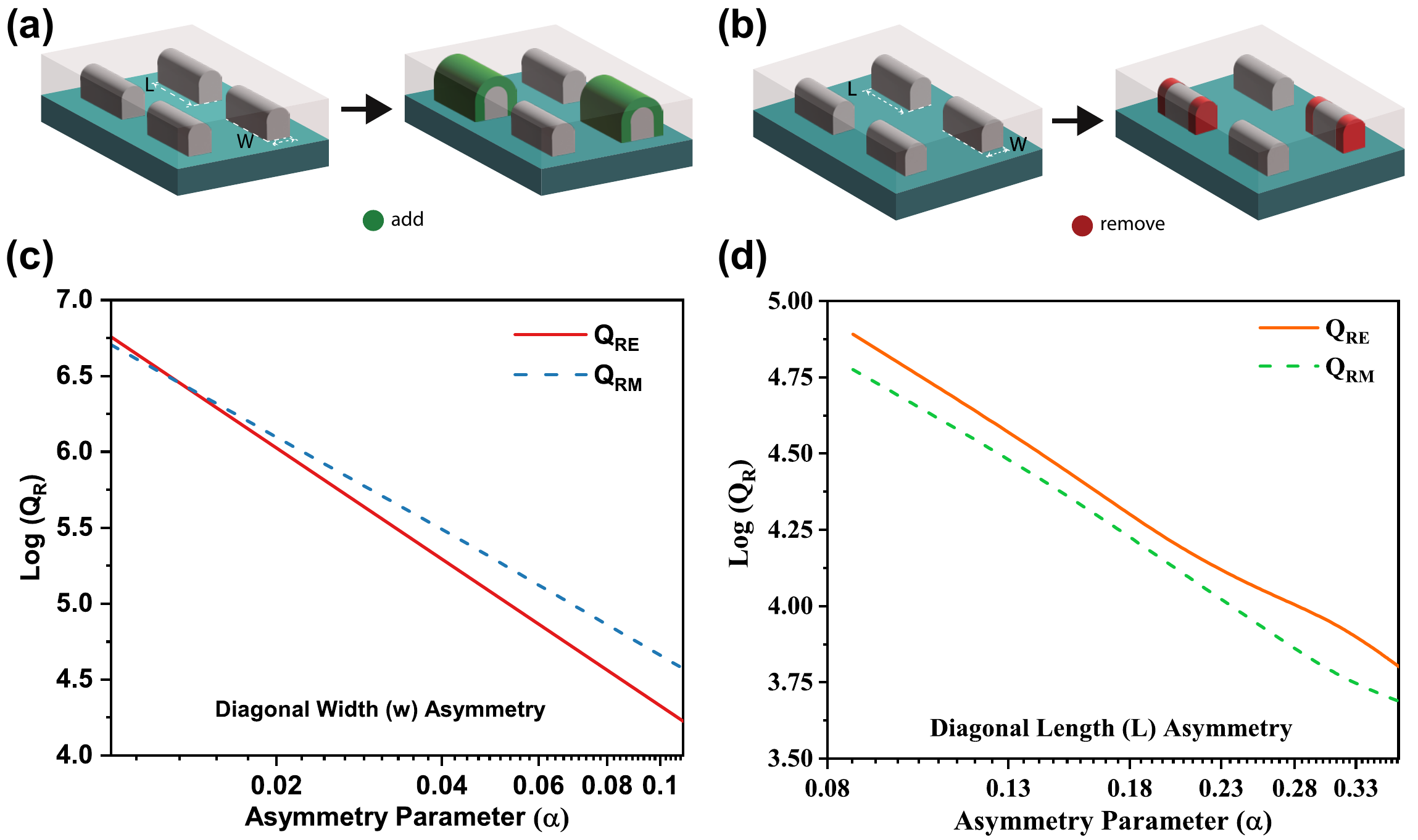}
        
    \caption{(a-b) 3D schematic of the MS with diagonal width and length asymmetry of the silicon nanobars, respectively. Red and green surrounding the area indicate the added and removed perturbation, respectively.   (c-d) Radiative Q-factor in logarithm scale of width and length asymmetry for varying asymmetric parameter ($\alpha$). Here, Q\textsubscript{RE} and Q\textsubscript{EM} refer to the radiative Q factor for ED-QBIC and MD-QBIC, respectively. }
    \label{fig:asym}
\end{figure*}

Two types of additional asymmetry were analyzed: the diagonal width asymmetry and the diagonal length asymmetry which are illustrated in Fig. \ref{fig:asym}(a) and Fig. \ref{fig:asym}(b), respectively. For diagonal width asymmetry, the symmetric width of the four domes was taken as 140 nm initially and then the widths of diagonally placed domes were changed from 142 nm to 164 nm to break the symmetry of the meta-atoms.

\noindent For the diagonal length asymmetry, the lengths of the diagonally placed domes varied from 175 nm to 185 nm. Here, the asymmetry parameter, $\alpha$ was defined by $\mathrm{\Delta L/L}$ and $\mathrm{\Delta W/W}$. Figure \ref{fig:asym}(c) and Fig. \ref{fig:asym}(d) show the radiative Q-factors for both ED-qBIC and MD-qBIC of diagonal width and diagonal length asymmetry. From the figure, we can see that Q-factors are almost proportional to $\mathrm{{\alpha}^{-2}}$ for both MD-qBIC and ED-qBIC resonances.

\section{Conclusion}
In conclusion, we propose an asymmetric dome-shaped dielectric metasurface to explore the different properties of BIC. We discovered two dips that correspond to the ED-qBIC and MD-qBIC in the transmission spectra of in-plane symmetry-broken MS in the near-IR region. The behavior of these qBICs was investigated in great detail, with particular focus placed on asymmetry characteristics such as and period along the y-axis, P\textsubscript{y}. We detected two distinct types of transmission spectrum behavior as a consequence of P\textsubscript{y} variation based on the superstate material refractive index: crossing and avoided-crossing of both qBIC resonances. Later on, we discussed two potential applications of the suggested MS: third harmonic generation and RI sensing and we reported that both applications can be implemented using the MS.  In the end, we demonstrated two different types of asymmetry analyses, namely diagonal length asymmetry and diagonal width asymmetry. We feel that MS design with these forms of asymmetries may have the potential in the future to be investigated further for a variety of applications.

\section{Acknowledgement} 
The authors appreciate Fawjia Shakhi from the Department of Electrical and Electronic Engineering, Shahjalal University of Science and Technology for the design-related illustration.

\section{Disclosures}
The authors declare no conflicts of interest.

\section{Data availability} Data underlying the results presented in this paper are not publicly available at this time but may be obtained from the authors upon reasonable request.

{\scriptsize
\bibliography{bibliography.bib}
\bibliographystyle{unsrt}
}

\end{document}